\documentclass[reprint,
superscriptaddress,
 amsmath,amssymb,
 aps,
prl,
floatfix,
]{revtex4-1}

\usepackage{physics}

\usepackage{graphicx}
\graphicspath{ {pdffig/}, {supfig/}, {images/}, {figs/} }
\usepackage{dcolumn}
\usepackage{bm}
\usepackage{hyperref}
\usepackage[mathlines]{lineno}
\usepackage{pifont}
\usepackage{xcolor}
\hypersetup{
	colorlinks,
	linkcolor={red!90!orange},
	citecolor={blue!90!black},
	urlcolor={pink!60!red}  
}
\usepackage[caption=false]{subfig}
\makeatletter
\newcommand{\colorcaption}[2][]{%
  \begingroup%
  \renewcommand{\@caption@fignum@sep}{ (color online). }%
  \caption[#1]{#2}%
  \endgroup%
}
\makeatother

\begin{document}

\title{Line nodes, Dirac points and Lifshitz transition in 2D nonsymmorphic photonic crystals}

\author{Jun Yu Lin}\thanks{These authors contributed equally to this work.}
\affiliation{Department of Physics, Sun Yat-sen University, SYSU Guangzhou Campus, Guangzhou 510275, China}

\author{Nai Chao Hu}\thanks{These authors contributed equally to this work.}
\affiliation{Department of Physics, Sun Yat-sen University, SYSU Guangzhou Campus, Guangzhou 510275, China}

\author{You Jian Chen}
\affiliation{Department of Physics, Sun Yat-sen University, SYSU Guangzhou Campus, Guangzhou 510275, China}
\author{Ching Hua Lee}
\email{calvin-lee@ihpc.a-star.edu.sg}
\affiliation{Institute of High Performance Computing, A *STAR, 138632, Singapore}
\author{Xiao Zhang}
\email{zhangxiao@mail.sysu.edu.cn}
\affiliation{Department of Physics, Sun Yat-sen University, SYSU Guangzhou Campus, Guangzhou 510275, China}

\date{\today}

\begin{abstract}
Topological phase transitions, which have fascinated generations of physicists, are always demarcated by gap closures. In this work, we propose very simple 2D photonic crystal lattices with gap closure points, i.e. band degeneracies protected by nonsymmorphic symmetry. Our photonic structures are relatively easy to fabricate, consisting of two inequivalent dielectric cylinders per unit cell. Along high symmetry directions, they exhibit line degeneracies protected by glide reflection symmetry, which we explicitly demonstrate for $pg,pmg,pgg$ and $p4g$ nonsymmorphic groups. In the presence of time reversal symmetry, they also exhibit point degeneracies (Dirac points) protected by a $Z_2$ topological number associated with crystalline symmetry. Strikingly, the robust protection of $pg$-symmetry allows a Lifshitz transition to a type II Dirac cone across a wide range of experimentally accessible parameters, thus providing a convenient route for realizing anomalous refraction. Further potential applications include a stoplight device based on electrically induced strain that dynamically switches the lattice symmetry from $pgg$ to the higher $p4g$ symmetry. This controls the coalescence of Dirac points and hence the group velocity within the crystal.
\end{abstract}

\maketitle


\textit{Introduction--}
In the recent few years, there has been considerable interest in the search for novel degeneracies associated with nonsymmorphic symmetries in electronic structures\cite{2016arXiv160303093B,Wang2016,kanedouble,muechler2016tilted,schoop2016dirac}. Since these degeneracies usually have topological origins, their study is a natural extension of the larger program of discovering new topological phases
, both theoretically and experimentally\cite{RevModPhys.83.1057, RevModPhys.82.3045, moore2010birth, chen2009experimental, zhang2009topological,RevModPhys.83.1057, RevModPhys.82.3045, moore2010birth, chen2009experimental, zhang2009topological,yu2010quantized, chang2013experimental, PhysRevLett.95.146802, bernevig2006quantum,PhysRevLett.50.1395, PhysRevLett.48.1559, Sun2011NoLandau, Xu2016, He2016,lee2017band}.
Such phases, which are protected by symmetry and/or a nontrivial topological index, possess interesting physically manifestations like boundary states, quantized response or exotic quasi-particle excitations. While a large number of topological phases have been theoretically identified and classified for different symmetry classes and dimensions\cite{Schnyder2008, Kitaev:2009mg, MHua2016}, only a handful have been experimentally realized in electronic systems. This is fundamentally due to the limited tunability of the Fermi level and atomic configurations.

Hence the push towards the realization of topological phases in alternative, artificial systems like photonic\cite{haldane2008possible, PhysRevA.78.033834, wang2008reflection, Dong2016,soskin2016singular,goryachev2016reconfigurable}, phononic\cite{salerno2014dynamical,nash2015topological,wang2015topological,susstrunk2015observation,yang2015topological,zhu2015topologically,fleury2015floquet,paulose2015selective,ong2016transport,huber2016topological,liu2016topological,lu2016observation,lee2017dynamically} and cold atom\cite{sun2012topological,furukawa2015excitation,qin2016topological} systems both in 2D and 3D\cite{Parameswaran2013, PhysRevB.90.085304, PhysRevB.93.045429, Lu:15, lu2016symmetry, 2016arXiv160701862K}, where topological invariants can be defined in analogy to those in conventional electronic systems. Photonic systems are particularly convenient for probing novel topological physics\cite{PhysRevLett.110.076403, xiao2015,2016arXiv160702918C,He03052016} due to their exactly solvable governing equations and lack of fundamental length scale\cite{joannopoulos2011photonic}. Indeed, topological phases have been discovered in various photonic systems with different symmorphic symmetries. Nontrivial edge modes have been observed in two-dimensional (2D) photonic crystals with $C_4$\cite{WenXiao-Gang}, $C_6$\cite{PhysRevLett.114.223901} or mirror symmetry\cite{chen2014experimental}, and topological Weyl points and nodel lines have been found in three-dimensional (3D) photonic crystals with gyroid structures or screw symmetry\cite{lu2013weyl,type2}. One important 
advantage of photonic crystals is that they can contain features of any desired shape, i.e. an ellipsoid, which is impossible to realize in electronic systems. This additional freedom shall play a crucial role in our implementation of 2D nonsymmorphic symmetry groups. 

Motivated by the richness of nonsymmorphic symmetry, we present specially designed 2D photonic crystals symmetric under the four nonsymmorphic wallpaper groups $pg$, $pmg$, $p4g$ and $pgg$. Compared to previous proposals involving strongly spin-orbit coupled systems\cite{kane2015dirac}, our lattice structures are extremely simple and amenable to experimental realization, consisting of only two inequivalent elliptical dielectric structures per unit cell. 
Various combinations of topologically robust Dirac points (DPs) and gapless line nodes exist depending on the nonsymmorphic symmetry group. 

Tuning our photonic crystals while preserving nonsymmorphic symmetry gives rise to various phenomena with potential technological applications. With $pg$ symmetry preserved, we show that a Lifshitz transition to a type-II DP occurs across a large range of realistic photonic rod shapes and dielectric constants. Analogous Lifshitz transitions have attracted considerable interest in the Weyl semimetal community\cite{soluyanov2015type,muechler2016tilted, wang2016mote,deng2016experimental,koepernik2016tairte}, and in our context leads to the phenomenon of anomalous refraction where an incident ray produces not one but two refracted rays. The sensitivity of the photonic bandstructure to the lattice symmetry brings forth the possibility of optical devices with mechanically-induced properties, as detailed in our stoplight device proposal.

Beginning with a pedagogical justification of the appearance of line degeneracies in a tight-binding (TB) Hamiltonian with $pg$ symmetry, we then extend the discussion to point degeneracies (Dirac points) emerging in higher orbitals. With the help of homotopy arguments, we analyze their topological properties, as well as numerically demonstrate their robustness through a protected $Z_2$ topological number. Finally, we detail the occurrence of Lifshitz transitions to type-II DPs, which are physically manifested through anomalous refraction. We conclude by proposing a stoplight device based on the C4-symmetry protected doubled DP.

\textit{TB model construction--}
A crystal with nonsymmorphic symmetry maps into itself under a combination of a point symmetry operation (i.e. reflection) and a fractional unit cell translation. 
Bandstructure degeneracies appear due to the existence of higher-dimensional projective representations of the nonsymmorphic symmetry group at certain momenta.

To understand the effect of nonsymmorphic symmetry, we first introduce the TB description of a photonic system\cite{albert2002photonic}. 
We focus on the case where waves propagate in an electric field parallel to the rod axis, forming the so-called harmonic transverse magnetic (TM) modes. The $n$-th mode $\vec E_{n,\mathbf{k}}=E_{n,\mathbf{k}}(\mathbf{r})\hat{z}$ obeys Maxwell's equation
\begin{equation}
    \nabla^{2}\vec E_{n,\mathbf{k}}(\mathbf{r})=-\frac{\omega^2_{n,\mathbf{k}}}{c^2}\varepsilon_p(\mathbf{r})\vec E_{n,\mathbf{k}}(\mathbf{r})
\end{equation}
where $\varepsilon_p(\mathbf{r})$ is the dielectric function of the periodic medium, and $\omega^2_{n,\mathbf{k}}$ is its frequency. If we rescale the modes via $\phi_{n,\mathbf{k}}(\mathbf{r})=\sqrt{\varepsilon_p(\mathbf{r})}E_{n,\mathbf{k}}(\mathbf{r})$, the eigenvalues of the Hermitian operator $H=-\frac{1}{\sqrt{\varepsilon_p(\mathbf{r})}}\nabla^{2}\frac{1}{\sqrt{\varepsilon_p(\mathbf{r})}}$ form
the effective band structure viz.
\begin{equation}
    H\phi_{n,\mathbf{k}}(\mathbf{r})=\frac{\omega^2_{n,\mathbf{k}}}{c^2}\phi_{n,\mathbf{k}}(\mathbf{r}).
		\label{TB0}
\end{equation}
This equation is the direct analog of the TB Schr\"{o}dinger's equation of an electronic system, where $\phi_{n,\mathbf{k}}$ represent its Bloch states. The hopping terms of this photonic TB Hamiltonian can be determined from the overlaps of the single dielectric rod eigenstates, which are well-localized like the orbitals of a single atom. Solving Eq.~\eqref{TB0}, we obtain the photonic band structure $\omega^2_{n,\mathbf{k}}/c^2$.

A nonsymmorphic crystal has at least two different components i.e. ``atoms'' in its unit cell, which are separated
by a non-primitive lattice vector. 
The effective TB Hamiltonian can be expressed in the normalized basis 
\begin{equation}
    \psi_{\alpha,\nu,\mathbf{k}}(\mathbf{r}) = 
		\sum_{\mathbf{R}}e^{i\mathbf{k}\cdot (\mathbf{R}+\mathbf{r}_{\alpha})}
    \varphi_\nu(\mathbf{r} -\mathbf{R} - \mathbf{r}_{\alpha}),
\end{equation}
in analogy to the Wannier basis for electronic systems\cite{vanderbilt,lee2013,lee2014lattice,chaoming}. Here $\varphi_\nu(\mathbf{r} -\mathbf{R} - \mathbf{r}_{\alpha})$  are the L\"{o}wdin orbitals representing the TM modes, where
$\mathbf{R}$ is the usual lattice vector, $\mathbf{r}_{\alpha}$ the position vector of atom $\alpha$, and $\nu$ its orbital degree of freedom. 

Due to the non-primitive lattice vector, the basis obeys extra constraints in addition to the Bloch condition: We have $\psi_{\alpha,\nu,\mathbf{k}+\mathbf{G}}(\mathbf{r}) = e^{i\mathbf{G}\cdot\mathbf{r}_{\alpha}} \psi_{\alpha,\nu,\mathbf{k}}(\mathbf{r})$,
where $\mathbf{G}$ is any reciprocal lattice vector. Hence the off-diagonal TB Hamiltonian matrix elements defined by $H_{\alpha\beta}(\mathbf{k}) = \int d\mathbf{r}\psi_{\alpha,\nu,\mathbf{k}}^*(\mathbf{r})\hat{H}\psi_{\beta,\nu,\mathbf{k}}(\mathbf{r})$ are periodic only up to a phase (i.e. form a projective representation of lattice translation):
\begin{equation}
H_{\alpha\beta}(\mathbf{k}+\mathbf{G}) = e^{i\mathbf{G}\cdot\mathbf{r}_{\Delta}}H_{\alpha\beta}(\mathbf{k}), 
\end{equation}
where $\mathbf{r}_{\Delta} = \mathbf{r}_{\beta}-\mathbf{r}_{\alpha}$. 
Lattices with different nonsymmorphic symmetries can be achieved with photonic cavities of different positions and orientations. Here, we shall implement them using dielectrics shaped as elliptical cylinders.
\begin{figure}[!t]
    \centering
    \includegraphics[width=.96\linewidth]{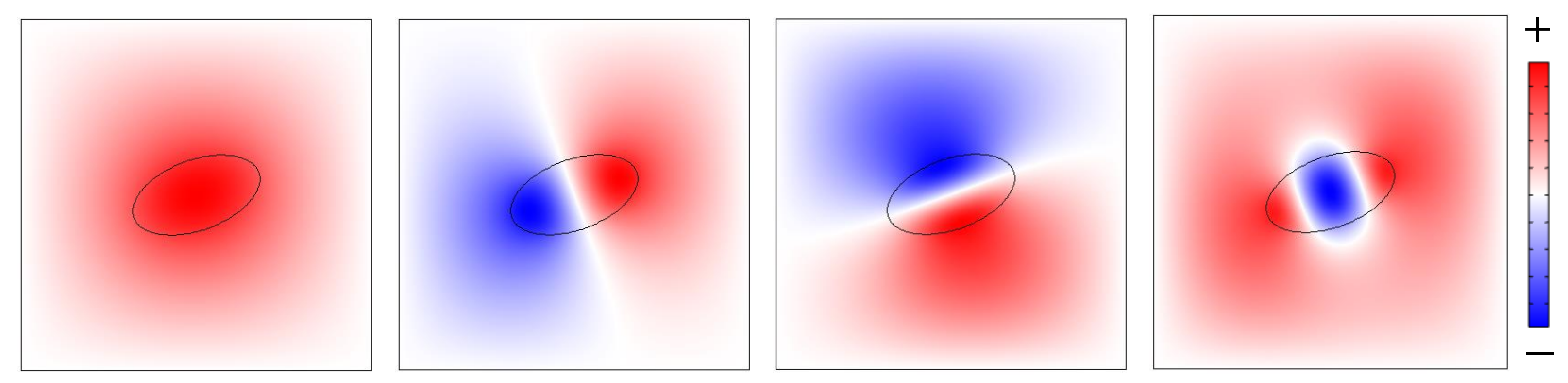}%
\caption{The $s$, $p_x$, $p_y$ and $2s$ orbitals above each elliptical cylinder in the photonic lattice unit cell. Note that they are neither isotropic nor aligned with the $x$ and $y$ axes, like the ellipses themselves.}
\label{fig:orbital}
\end{figure}

\begin{figure}[!t]
\subfloat[]{\label{fig:pg}%
  \includegraphics[width=0.23\textwidth]{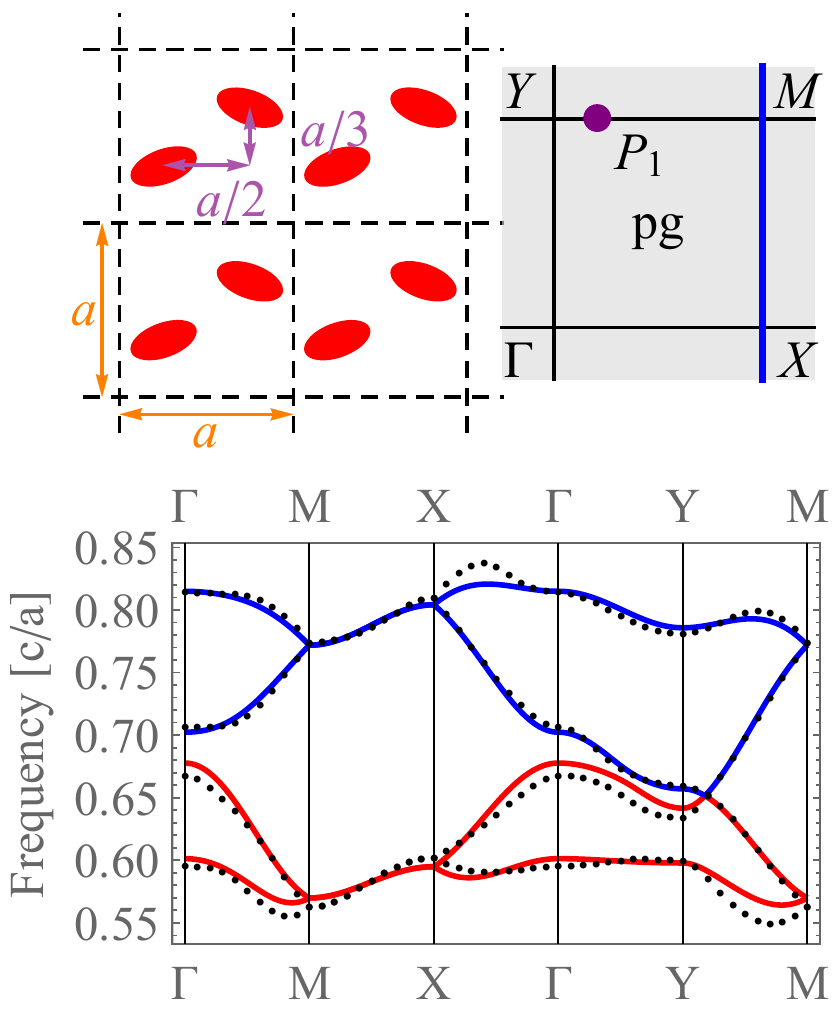}%
}
\subfloat[]{\label{fig:pmg}%
  \includegraphics[width=0.23\textwidth]{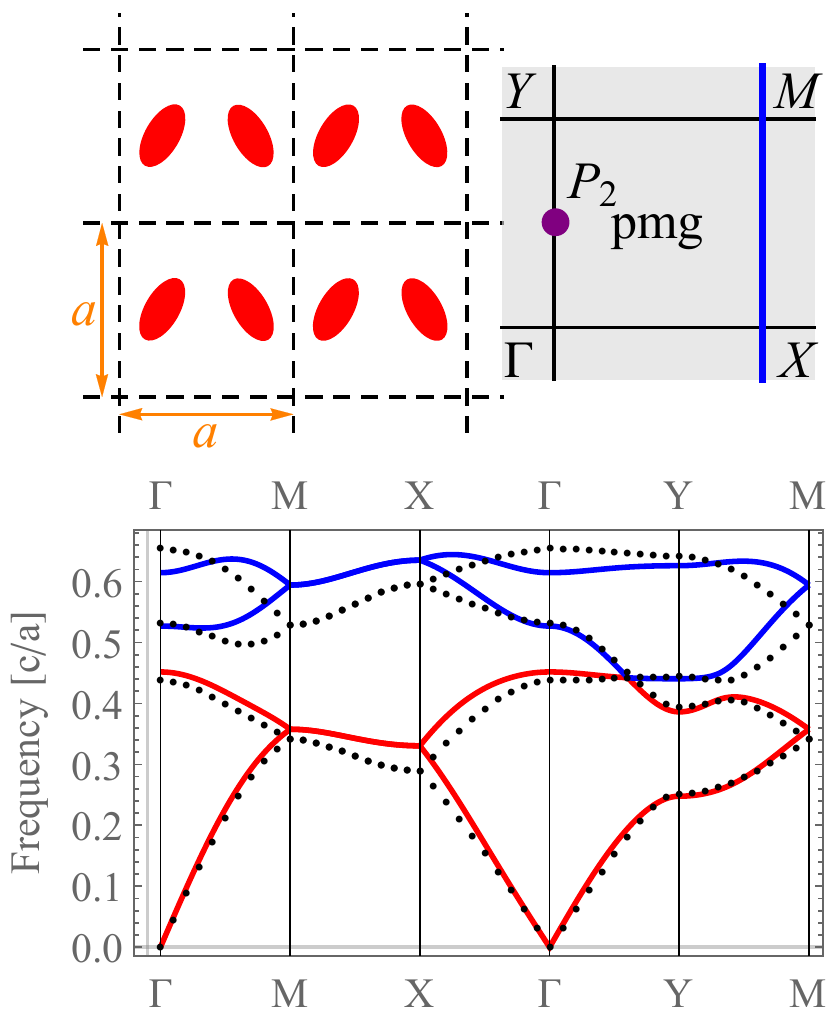}%
}\\
\subfloat[]{\label{fig:p4g}%
  \includegraphics[width=0.23\textwidth]{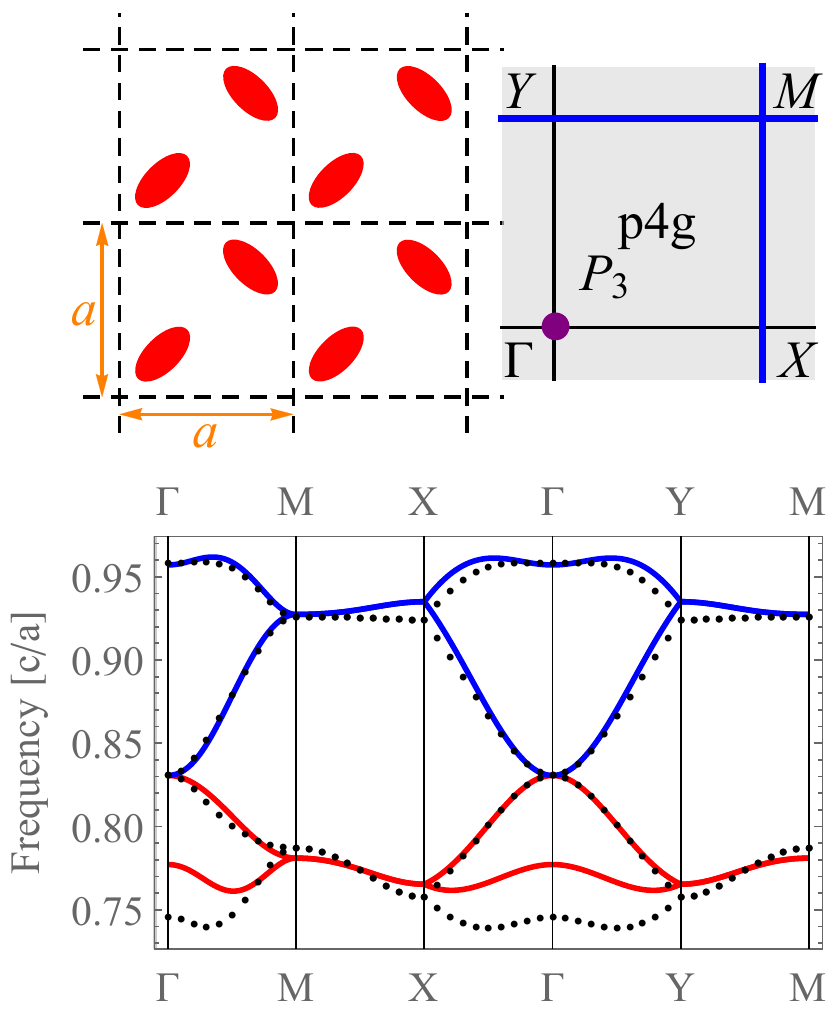}%
}
\subfloat[]{\label{fig:pgg}%
  \includegraphics[width=0.23\textwidth]{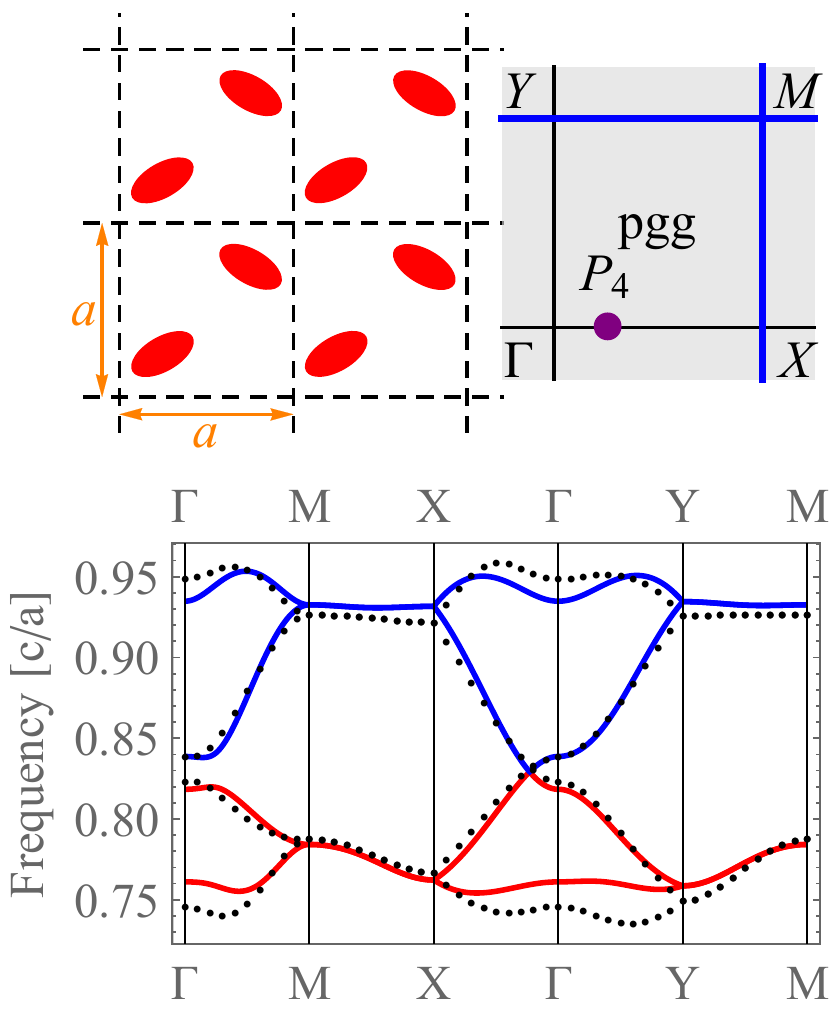}%
}
\caption{
The lattice structure, nodal positions and band dispersions for the four lattices we considered, with $pg$, $pmg$, $p4g$ and $pgg$ symmetry respectively (a to d). Results from our effective TB model (colored lines, see Supp. Materials~\cite{Note1}) agree closely with COMSOL simulation results (black dashed lines).  
(a) The $pg$ bands from the $p_x$ and $p_y$ orbitals. There exists a Dirac point (DP) $P_1$ along $Y$-$M$ and a nodal line along $M$-$X$ protected by $\{m_y|\tau_x\}$.
(b) By rotating the ellipses in the $pg$ lattice to respect an additional mirror symmetry $m_x$, we obtain the $pmg$ lattice. The nodal line remains unchanged but the DP $P_2$ is now along $\Gamma$-$Y$.
(c) The lattice with $p4g$ symmetry, which has mirror symmetry along two diagonals and an additional glide reflection symmetry $\{m_x|\tau_y\}$. 
The nodal lines persist but there is now a doubly degenerate DP $P_3$ at the $\Gamma$ point.
(d) The $pgg$ lattice obtained by breaking the $C_4$ symmetry of the $p4g$ lattice through arbitarry rotation of the ellipses. 
 The previously doubly degenerate DP decomposes into two singly degenerate DPs between $\Gamma$ and $\pm X$, one of which is visible here.
} 
\label{fig:band}
\end{figure}

\textit{Line nodes from nonsymmorphic symmetry--}
As a first illustration of how nonsymmorphic symmetry can lead to degeneracies, consider the lowest two bands of the photonic crystal. These two bands correspond to the two $|s\rangle$ orbitals above the two inequivalent elliptical dielectric regions (labeled as $A$ and $B$). 
Note that the $\ket{s}$ orbitals are not isotropic due to the anisotropy of the elliptical cylinder. From Fig.~\ref{fig:pg}, we see that the photonic crystal (PhC) has the symmetry of one of the simplest nonsymmorphic group $pg$, which only contains glide reflections. 
The glide reflection operators are conventionally denoted by $g_y = \{m_y|\tau_x\}$, where  $m_y \psi(x,y,z)=\psi(x,-y,z)$ and $\tau_x \psi(x,y,z)=\psi(x+\frac{a}{2},y,z)$. 
Denoting orbital overlaps by $J^{AB}_{x,y}$, an immediate consequence of this glide reflection symmetry is that $J^{BA}_{x,y}=J^{AB}_{x,y}=J^{AB}_{-x,y}$ and $J^{AA}_{x,y}=J^{BB}_{x,y}$. Hence $H_{AB}(\pi/a,k_y)=0$ and $H_{AA}(\pi/a,k_y)=H_{BB}(\pi/a,k_y)$, i.e. we have a degenerate line node along $k_x=\pi/a$ (line $MX$). Analogous arguments hold for generic line degeneracies at the BZ boundary (see Fig.~\ref{fig:band} for more examples.)


\textit{Protected Dirac points--}
Besides protecting line nodes, nonsymmorphic symmetry also protects Dirac crossings in the photonic bandstructure by protecting the $Z_2$ topological number of the 1D Berry phase\cite{kariyado2013symmetry}. As detailed in the Supplement\cite{Note1}, point degenaracies must exist at the $Z_2$ jumps. By adjusting the relative positions and orientations of the cylinders in our photonic crystal, various nonsymmorphic symmetries $pg$, $pmg$, $pgg$ and $p4g$ (Fig.~\ref{fig:band}) can be implemented, each giving rise to protected Dirac points in certain bands. One observes the splitting, fusion and motion of these DPs as the cylinders are continuously modified.

Perturbing the 4-orbital TB model around each degeneracy\cite{sun2012topological,Note1} yields an effective 2-band Hamiltonian $H_{\mathbf{k}} = \mathbf{h}_{\mathbf{k}} \cdot \bm{\sigma}$ characterized by the $ \mathbf{h}_{\mathbf{k}}$ vector, where $\bm{\sigma}$ are the Pauli matrices. When sublattice symmetry is respected, as in the $pmg$, $pgg$ and $p4g$ symmetry groups, $\mathbf{h}_{\mathbf{k}}$ is confined to a plane and a winding number $w$ can be defined for the mapping $\mathbf{h}_{\mathbf{k}}$: $\mathbb{S}^1 \mapsto \mathbb{S}^1$ along a closed loop around the gapless point:
\begin{equation}
w = \oint\frac{d\mathbf{k}}{2\pi}\left[\frac{h_1}{|\mathbf{h}|}\nabla\frac{h_2}{|\mathbf{h}|}-\frac{h_2}{|\mathbf{h}|}\nabla\frac{h_1}{|\mathbf{h}|}\right] \in \mathbb{Z}.
\label{winding}
\end{equation}
To elucidate our findings in more detail:

\paragraph{pg group:} 
We consider the same lattice as before (Fig.~\ref{fig:pg}) 
, but now focus on the four bands spanned by orbitals $\ket{A,p_x}$, $\ket{B,p_x}$, $\ket{A,p_y}$ and $\ket{B,p_y}$. 
A Dirac point $P_1$ exists along $Y$-$M$, with gap opening up if time reversal or nonsymmorphic $pg$ symmetry is broken, i.e. by using a distorted magneto-optical dielectric. Along MX, a line node exists for reasons explained earlier.

\begin{figure}[t]
    \centering
    \includegraphics[width=\linewidth]{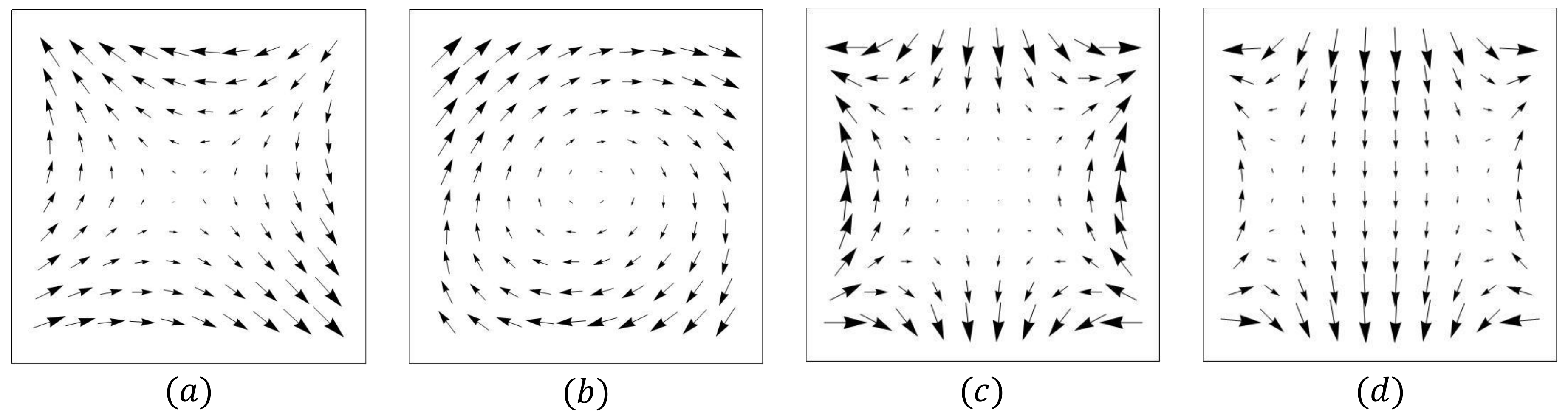}
    \caption{The configurations of $\mathbf{h}$ near the DPs of lattices with $pmg$, $p4g$ and $pgg$ symmetries. (a) and (b) depict the two DPs ($ \pm P_2$ between $\Gamma$ to $\pm Y$) of the $pmg$ lattice, with windings $w = -1$ and $w = 1$. 
    (c) The $w=-2$ $\mathbf{h}$ around the quadratic degeneracy ($P_3$) at $\Gamma$ for the $p4g$ symmetric lattice.
    (d) With $p4g$ broken to $pgg$, the above $w=-2$ degeneracy splits into to two DPs ($\pm  P_4$) along $-X$ to $X$, each with winding $w=-1$.
    }
    \label{fig:hvector}
\end{figure}

\paragraph{pmg group:} Besides glide reflection symmetry as in $pg$, the lattice also contains an additional mirror symmetry (Fig.~\ref{fig:pmg}). Within the bands spanned by $\ket{A,s}$, $\ket{B,s}$, $\ket{A,p_y}$ and $\ket{B,p_y}$, Dirac cones $\pm P_2$ between $\Gamma$ and $\pm Y$ appear without fine-tuning. Here, it is the mirror symmetry subgroup of $pmg$ that is essential in protecting the Dirac crossing. 
By contrast, the nodal line requires the symmetry under the glide operation.

\paragraph{p4g group:} $p4g$ symmetry consists of mirror symmetries along the two diagonals and glide reflection symmetries $g_x=\{m_x |\tau_y\}$ and $g_y=\{m_y |\tau_x\}$ (Fig. 2(c)). 
Due to the extra $C_4$ rotational symmetry, the $\Gamma$ point hosts a quadratically degenerate point $P_3$ in the space of orbitals $\ket{A,p_y}$, $\ket{B,p_y}$, $\ket{A,2s}$ and $\ket{B,2s}$, with a $\mathbf{h}_{p4g}$ winding of $w=-2$ (Fig.~\ref{fig:hvector}(c)). 
To understand exactly which symmetry subgroup is necessary for protecting this double degeneracy, we proceed to break the $C_4$ rotational symmetry next.

\paragraph{pgg group:} Upon breaking $C_4$ rotation symmetry by rotating each of the elliptical cylinders, we obtain the $pgg$ lattice  (Fig.2(d)) from the previous $p4g$ lattice (Fig.2(c)). 
The quadratic degeneracy at $\Gamma$ decomposes into two linear ($w=-1$) DPs $P_4$ located either along $\Gamma$-$ X$, $\Gamma$-$ Y$ or their mirror inverses, depending on how the $C_4$ symmetry was broken. For the case of $\Gamma$-$X$ shown, $P_4$ is gapped by breaking both $g_x$ and $C_2$ (but not $g_y$). Hence either $g_x$ and parity symmetry can protect $P_4$, but only $g_x$ can confine $P_4$ along $\Gamma$-$ X$.

\begin{figure}[!t]
\subfloat[]{\label{fig:type2lat}%
  \includegraphics[width=0.15\textwidth]{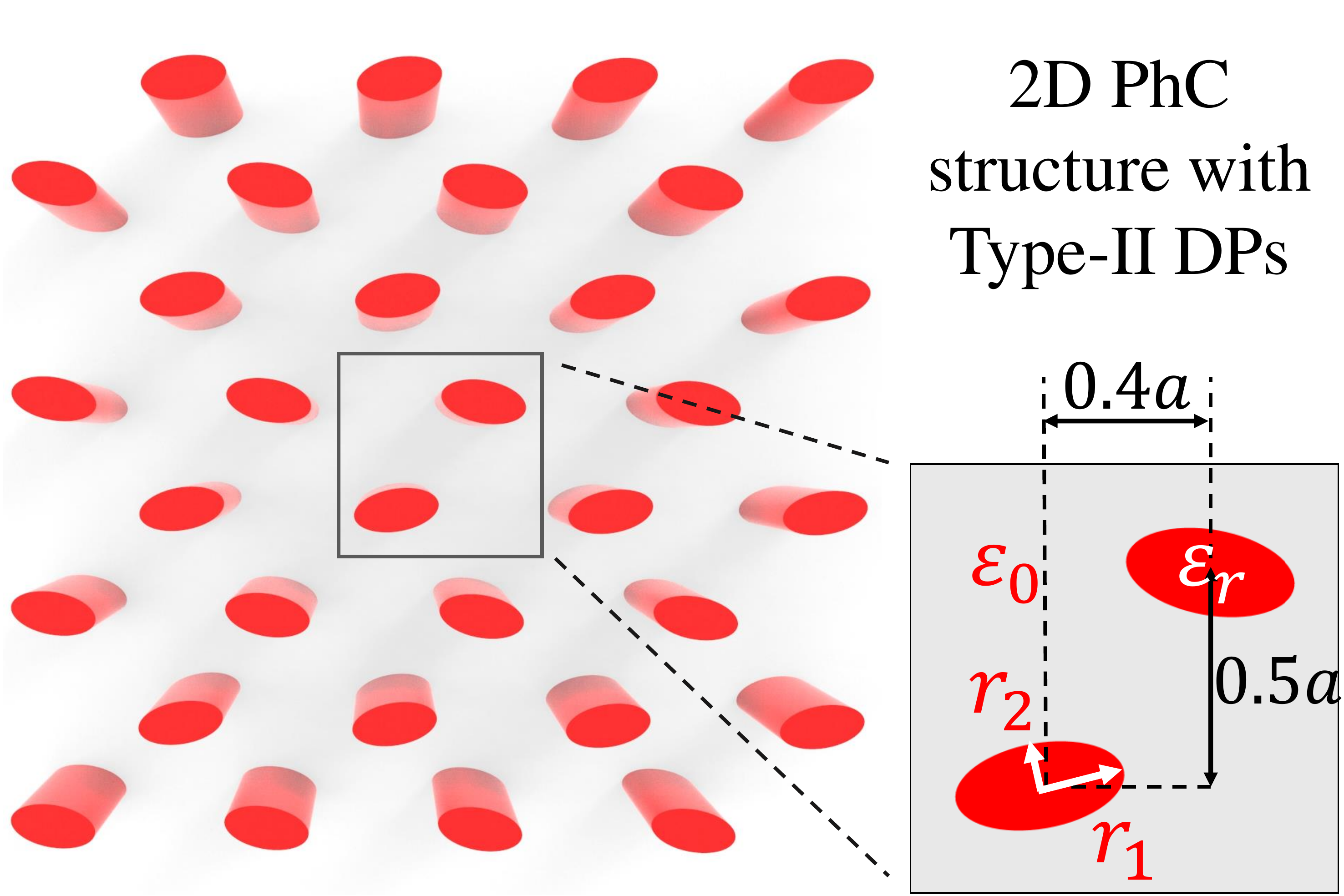}%
}
\subfloat[]{\label{fig:secondtype}%
  \includegraphics[width=0.15\textwidth]{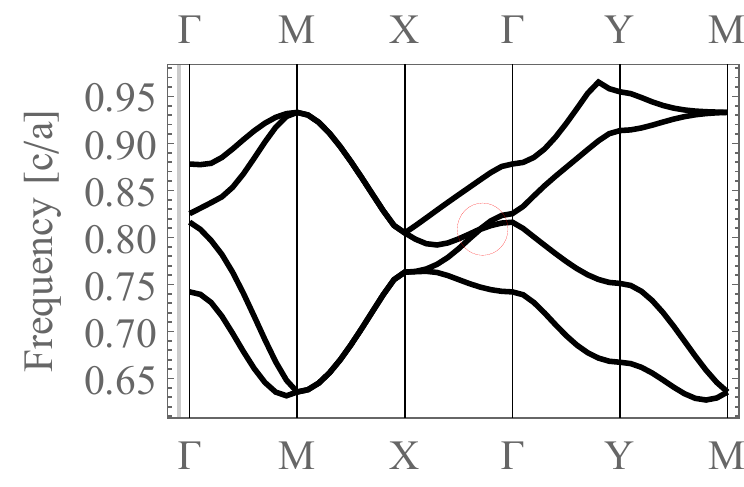}%
}
\subfloat[]{\label{fig:phase}%
  \includegraphics[width=0.17\textwidth]{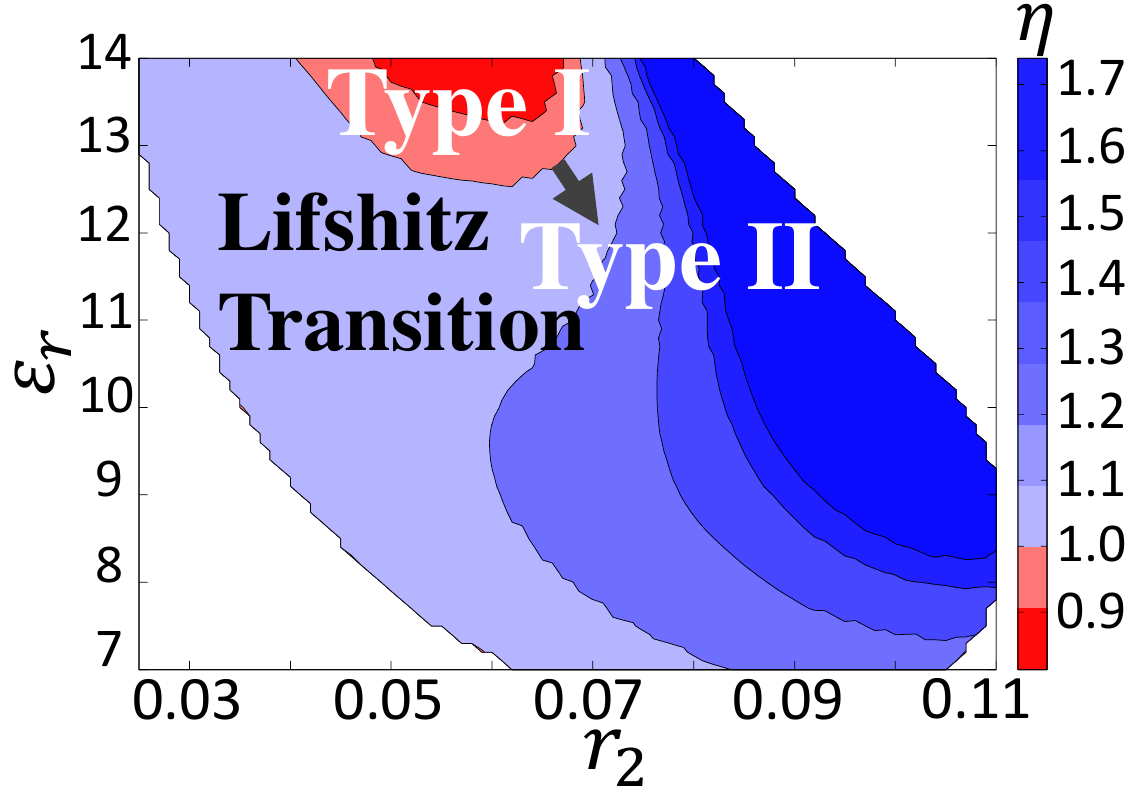}%
}
\caption{ The $pg$ lattice (a) and bandstructure (b) hosting type-II (tilted over) Dirac points. (c) Phase diagram for Lifshitz transition dependence on dielectric constant $\epsilon_r$ and aspect ratio $r_2/r_1$ of the ellipses, with $r_1=0.16a$ and orientation angles $\pm 80^\circ$. Type-I/II regions are marked in red/blue, while no DP exists in the white regions.
}
\label{fig:latticeandphase}
\end{figure}

\begin{figure}[t]
\subfloat[]{\label{fig:pgII}%
  \includegraphics[width=0.23\textwidth]{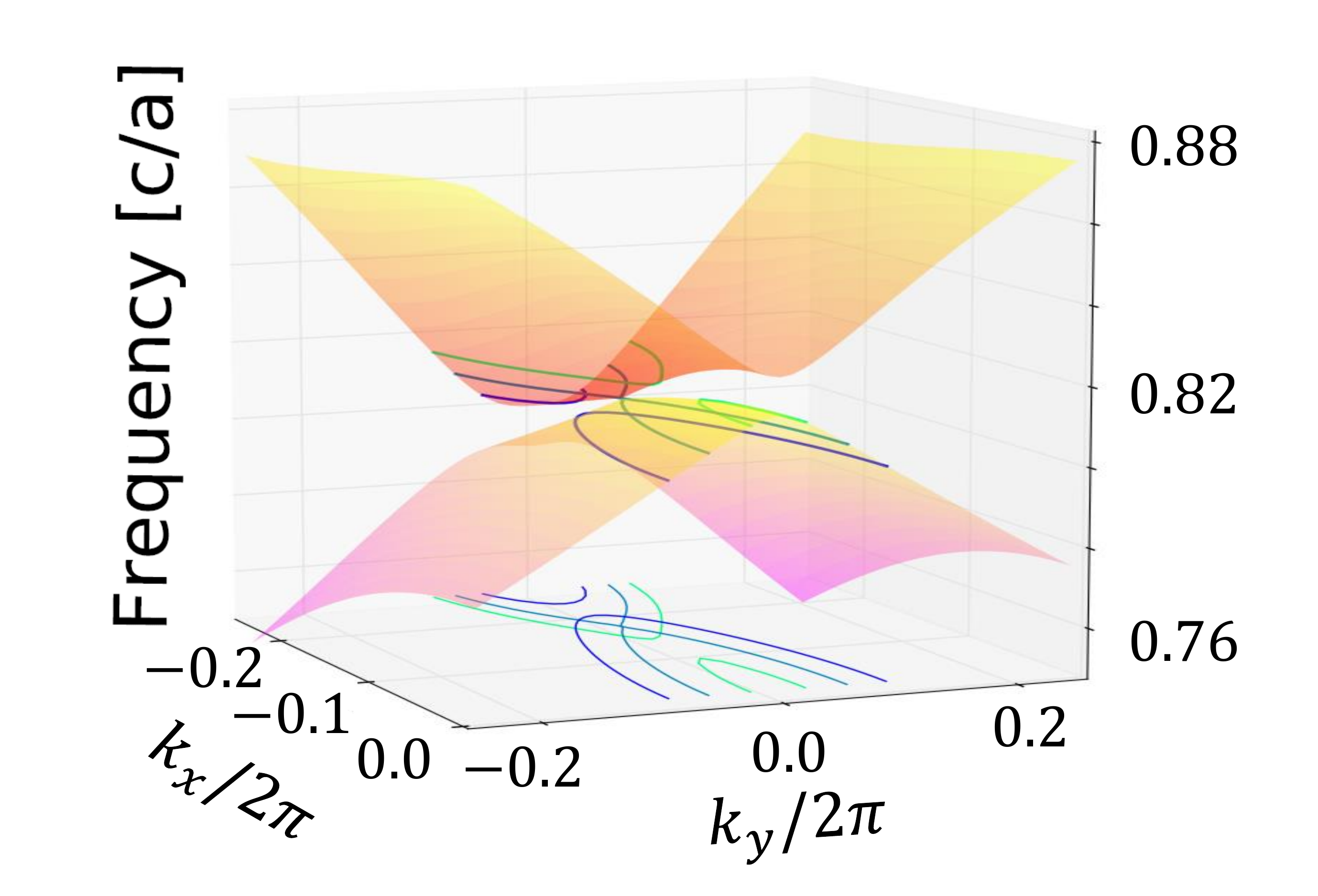}%
}
\subfloat[]{\label{fig:typeIIrefract}%
  \includegraphics[width=0.23\textwidth]{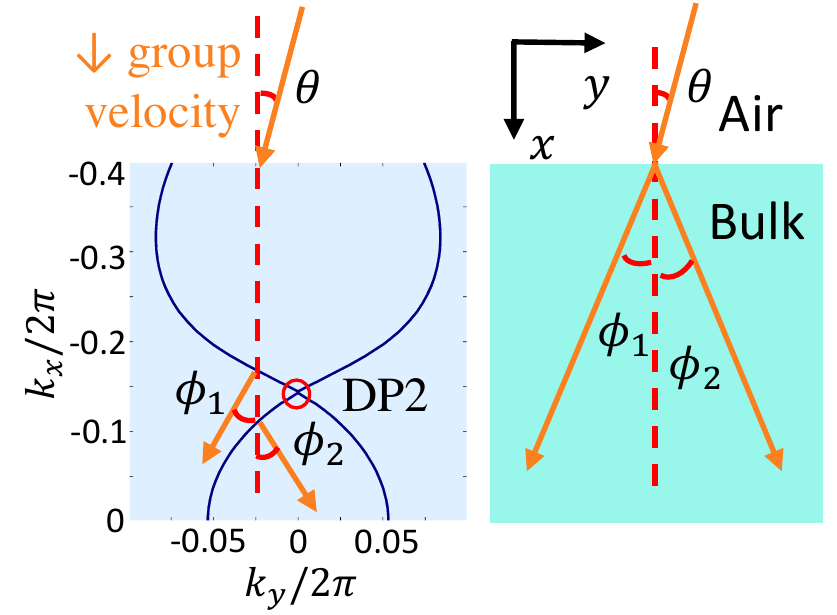}%
}
\caption{(a) 3D frequency plot near a type-II Dirac point, with characteristic cross-shaped isofrequency lines due to its tilt.
(b) Anomalous refraction from a type-II DP. Left panel: The refraction angles are aligned with the refracted group velocities, which are determined by the normals of the isofrequency lines. These lines intersect with the incident wavevector at the original tangential wavevector component $k_y$.
}
\label{fig:type2}
\end{figure}

\textit{Lifshitz transition and anomalous refraction --} 
Interestingly, nonsymmorphic symmetry protects the point degeneracies so robustly that a Dirac cone can ``tilt over'' and still remain gapless upon  large parameter tuning. When a type-I (upright) Dirac cone tilts over into a type-II (tilted over) Dirac cone, the isofrequency ``Fermi'' surface undergoes a topological change known as a Lifshitz transition, from an isolated point to a pair of intersecting lines (Fig.~\ref{fig:pgII}). Its 3D analog has attracted considerable attention\cite{type2}, especially in the context of Weyl semimetals\cite{soluyanov2015type,muechler2016tilted, wang2016mote,deng2016experimental,koepernik2016tairte,type2}. In our PhCs, a Lifshitz transition can be induced across a wide range of   nonsymmorphic symmetry preserving deformations, particularly when the dielectric constant $\epsilon_r$ or aspect ratio of the ellipses are varied (Fig. \ref{fig:phase}).

Near a tilted Dirac point, the Hamiltonian generically assumes the form
\begin{equation}
\mathcal{H}_{II}(\delta \bold{k}) = v_x \delta k_x \sigma_x + v_y \delta k_y \sigma_y + (u_x\delta k_x+u_y \delta k_y) \mathbb{I},
\label{HII}
\end{equation}
where $\delta \bf{k}$ is the displacement from the DP. In our case, the linear $u_y$ term is forbidden by glide symmetry. The tilt $ \eta=u_x/v_x$ is controlled by the last term: $\eta=0$ for an untilted type-I DP, and $|\eta|>1$ for a type-II DP. 

From Eq. \ref{HII}, the isofrequency contour is given by
\begin{equation}
\delta\omega=\omega-\omega_0=\eta v_x \delta k_x \pm \sqrt{v_x^2 \delta k_x^2+v_y^2\delta k_y^2},
\end{equation}
where $\omega_0 = 0.811c/a$ is the frequency of the DP for our $pg$ lattice (Figs. \ref{fig:latticeandphase} and \ref{fig:type2}). Due to the unique double multiplicity of isofrequency lines near the DP, an incident light ray on the PhC will be anomalously separated into two refracted rays within the PhC. As derived in the Supp. Materials~\cite{Note1}, the two anomalous refraction angles corresponding to an incident angle $\theta$ are given by  
\begin{equation}
    \phi^{\pm} = \pm \tan^{-1} \frac{ v_y^2 |\delta k_y|}{\eta v_x(\delta\omega-\eta v_x \delta k_x) + v_x^2 \delta k_x},
\end{equation}
where $|\delta k_y| = \sqrt{(\delta\omega-\eta v_x \delta k_x)^2-v_x^2 \delta k_x^2}/v_y$ and $\delta k_x = \frac{\omega}{c}\sin\theta$. For frequencies near $\omega_0$, $|\phi^\pm|\approx \tan^{-1}\left[\frac{v_y}{v_x}\frac1{\sqrt{\eta^2-1}}\right]$, which suggests that anomalous refraction \emph{requires} $|\eta|>1$. This is contrasted with ordinary optical media where only one refracted ray is observed.

\textit{Stoplight device--} The sensitive dependence of the photonic dispersion on lattice symmetry also suggests a potential stoplight device application. When the symmetry is reduced from $p4g$ to $pgg$ by breaking $C_4$ rotation, 
the quadratic DP splits into two linearly dispersive DPs (Fig.~\ref{fig:p4g}). A simple representative Hamiltonian is 
\begin{equation}H_{split}(k_x,k_y,k_0)=t H_d(k_x,k_y,k_0)\circ H_d(k_x,k_y,-k_0)
\label{Hsplit}
\end{equation} 
where $\circ$ is the Hadamard product, $t$ is a constant, $k_0$ controls the splitting and $H_d(k_x,k_y,k_0) = (k_x-k_0)\sigma_x + k_y\sigma_y$ gives a single DP located at $k_0$. For an incident ray along the x-axis, Eq. \ref{Hsplit} gives a resultant group velocity $v_x^g = {\partial \omega}/{\partial k_x}|_{k_0} = 2t k_0 $.

\begin{figure}[t]
\subfloat[]{\label{fig:device}%
\includegraphics[width=0.26\textwidth]{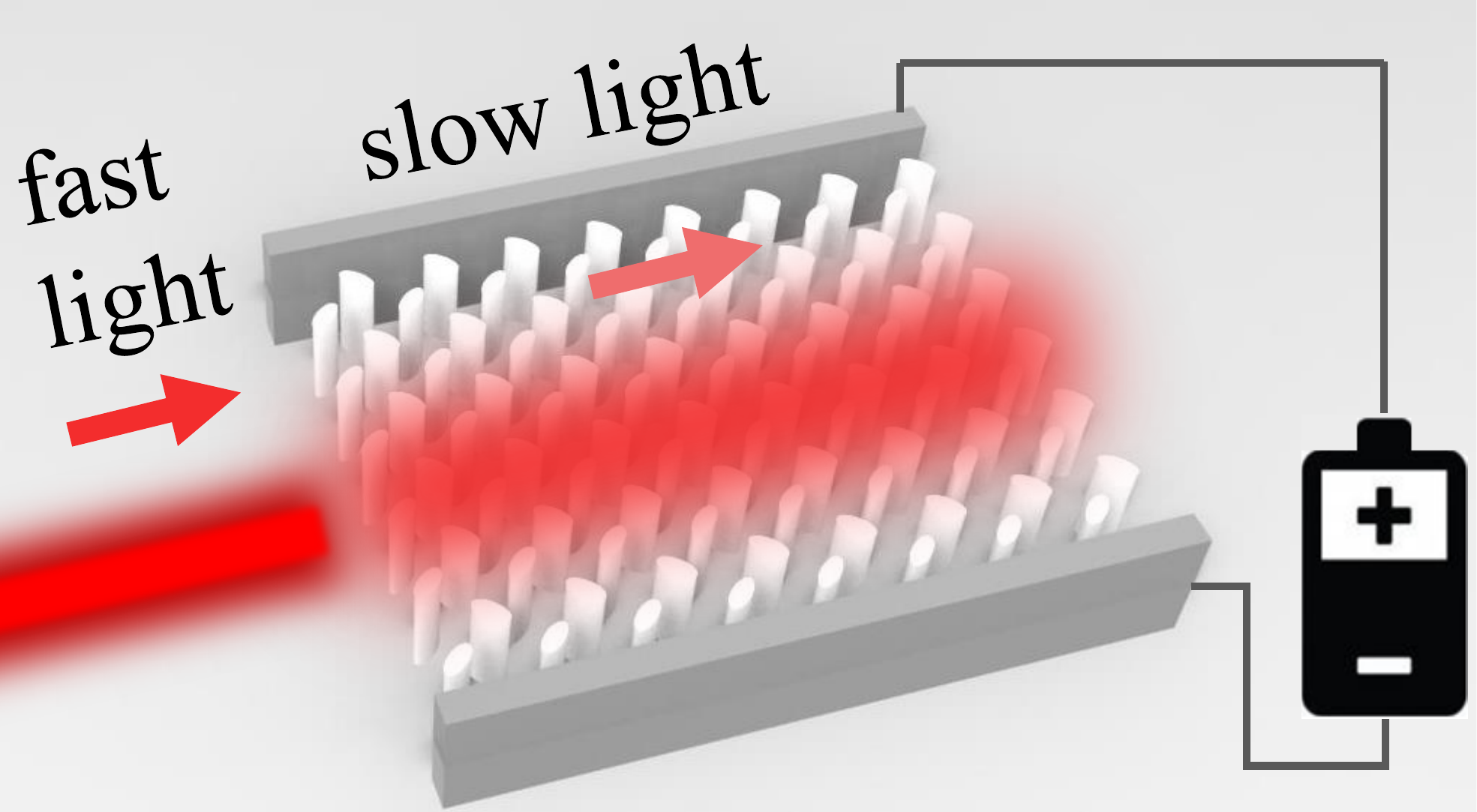}%
}
\subfloat[]{\label{fig:curve}%
  \includegraphics[width=0.21\textwidth]{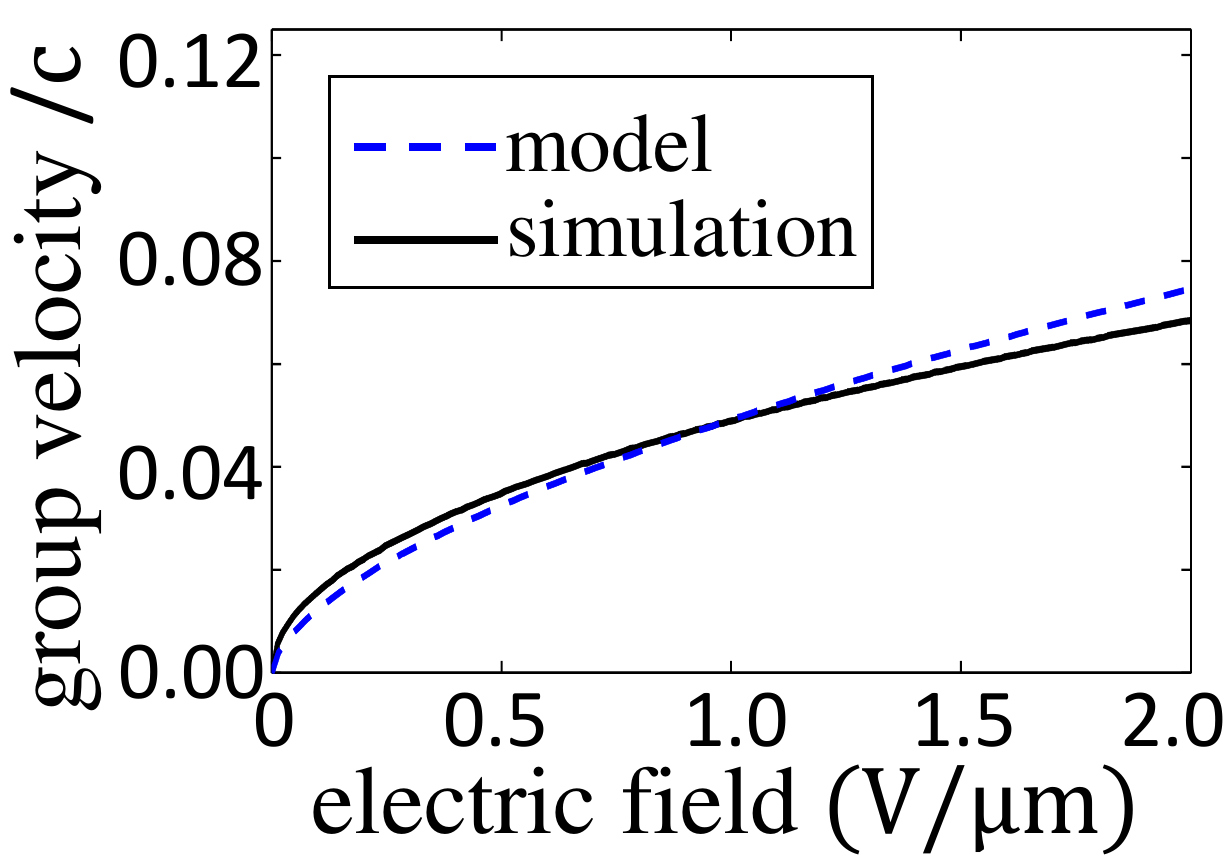}%
} 
    \caption{(a) Stoplight device with group velocity controlled by external applied voltage. An electric field deforms the silicone elastomer elliptic cylinders and breaks the $p4g$ symmetry to $pgg$, thereby modifying the dispersion. (b) The good agreement between COMSOL simulation results and the effective Hamiltonian in Eq. \ref{Hsplit}. 
		}
\end{figure}

The splitting $k_0$ can be dynamically controlled in suitable dielectrics exhibiting electric field induced strain, such as silicone elastomers\cite{pelrine2000high,kornbluh2000ultrahigh} with electrical permittivities within our desired range ($\epsilon_r = 11.82$)\cite{zhang2004}. As shown in Fig.~\ref{fig:device}, our proposed stoplight device consists of a $p4g$/$pgg$ symmetric PhC with elliptic cylinders made with 5\% 81-R hardener dissolved in silicone fluid blended with a 40\% copper-phthalocyanine oligomer. An appreciable strain of $12\%$ can be induced by a realistic applied electric field of $E \approx 25V/\mu m $ along the y-axis\cite{zhang2004}. 
With external applied electric field, the cylinders lengthen along the x-axis and shorten along the y-axis, breaking the symmetry from $p4g$ to $pgg$ and giving rise to nonzero $k_0$. Assuming negligible Poisson ratio, $k_0$ is empirically fitted to $k_0 = b E_a^D$ where $b = 0.16\mu m/aV, D = 0.605$ and $t = 0.162ac$, which agrees well with COMSOL simulation results as shown in Fig.~\ref{fig:curve}. Since it is easy to precisely control the applied voltage, our device will can function robustly as an optical switch that controls, or slows, light propagation significantly with precision. An added advantage is that since light propagation is stopped by $C_4$ symmetry, the elliptic cylinders may be replaced by cuboids or other $C_4$-symmetric shapes for  implementation convenience.   

\textit{Conclusion--}
In this work, we proposed very simple 2D photonic lattices whose bandstructures possess highly tunable line nodes and Dirac points protected by nonsymmorphic symmetry. Consisting of easily fabricated elliptical dielectric rods, these lattices can realize all the nonsymmorphic wallpaper symmetry groups $pg$, $pmg$, $p4g$ and $pgg$. 

Impressively, there exist large parameter regimes where the symmetry protected point degeneracy undergoes a Lifshitz transition into a type-II Dirac cone. Such exotic bandstructure topology have been associated with spectacular response properties\cite{soluyanov2015type}, and in our case results in anomalous refraction. The sensitivity of the bandstructure to lattice $p4g$ symmetry also leads to our proposal for a realistic light-stopping device.  

\begin{acknowledgments}
We thank Shanhui Fan, Quan Zhou, Pinaki Sengupta, Yuhan Liu, Qixian Liao and Guang-Jie Li for useful discussions.
\end{acknowledgments}
\bibliography{bib} 

\newpage

\onecolumngrid
\begin{widetext}
\textbf{\large Supplemental Online Material for ``Line nodes, Dirac points and Lifshitz transition in 2D nonsymmorphic photonic crystals" }\\[5pt]
\vspace{0.1cm}
{\small In this supplementary material, we detail: 
\begin{enumerate}
\item the parameters and tight binding models for our illustrative examples,
\item details of the $Z_2$ quantization of the 1D Berry phase due to nonsymmorphic symmetry,
\item details of the perturbation performed around the Dirac points,
\item the argument for line node protection under $pg$ symmetry, and
\item the derivation of anomalous refraction angles. 
\end{enumerate}}

\section{\label{sec:lns}Simulation parameters and tight binding Hamiltonians} 

The parameters for our simulations are given by Table.~\ref{tab:rep}.  
\begin{table}[ht!]
\centering
\caption{Lattice parameters for COMSOL simulation.}
\label{tab:rep}   
\begin{tabular}{ccccccc}
\hline 
group/type of DP & $r_1$ & $r_2$ & declination angle & position & $\varepsilon_r$ & figure\\
\hline
$pg$ Type-I & $0.19a$ & $0.1a$ & $10^\circ$ &(-a/4,-a/6) (a/4,a/6) & 14 & Fig.~\ref{fig:pg}\\
$pmg$ & $0.28a$ & $0.08a$ & $60^\circ$ &(-a/4,0) (a/4,0) & 10 & Fig.~\ref{fig:pmg}\\ 
$p4g$ & $0.19a$ & $0.1a$ & $45^\circ$  & (-a/4,-a/4) (a/4,a/4) & 14 & Fig.~\ref{fig:p4g}\\ 
$pgg$ & $0.19a$ & $0.1a$ & $40^\circ$&(-a/4,-a/4) (a/4,a/4) & 14 & Fig.~\ref{fig:pgg}\\
$pg$ Type-II & $0.16a$ & $0.1a$ & $-80^\circ$ &(a/4,a/5) (-a/4,-a/5) & 10 & Figs.~\ref{fig:secondtype},~\ref{fig:pgII}\\
$pg$ phase diagram & $0.16a$ & variable & $-80^\circ$ &(a/4,a/5) (-a/4,-a/5) & variable & Fig.~\ref{fig:phase}\\ 
\hline 
\end{tabular} 
\caption{$r_1$ and $r_2$ are the major and minor radii of the ellipses respectively, and the declination angle is the angle the major radius vector points below the horizontal.}
\end{table}

For theoretical analysis, we utilize the tight-binding description of the photonic lattices. The tight binding basis is spanned by $\phi_n = \sqrt{\epsilon(r)}E_n$, where $n$ can represent the $s$ orbital $s,p_x,p_y$ or $2s$ orbitals above each elliptical cylinder (Fig.~\ref{fig:orbital}). 

We write the 4$\times$4 TB Hamiltonian of Fig.~2 as:
\begin{align}
    H^{(4)} = \begin{pmatrix}
         H_{11} & H_{12} & H_{13} & H_{14} \\
         H_{21} & H_{22} & H_{23} & H_{24} \\
         H_{31} & H_{32} & H_{33} & H_{34} \\
         H_{41} & H_{42} & H_{43} & H_{44} \\
    \end{pmatrix}+ \epsilon_s \mathbb{I}_4
\end{align}
where $H_{ij}=H_{ji}^*$.
\paragraph{pg group:} Choosing $\ket{A,p_x}$, $\ket{B,p_x}$, $\ket{A,p_y}$ and $\ket{B,p_y}$ as the basis, the matrix elements of the 4$\times$4 Hamiltonian can be written as:
\begin{align}
H_{11} &=-\epsilon_0+s_1 \left(e^{-i k_x}+e^{i k_x}\right)+s_2 \left(e^{-i k_y}+e^{i k_y}\right)\\
\label{H11}
H_{12} &=t_2 \left(e^{-\frac{i k_x}{2}-\frac{2 i k_y}{3}}+e^{\frac{i k_x}{2}-\frac{2 i k_y}{3}}\right)-t_1 \left(e^{\frac{i k_y}{3}-\frac{i k_x}{2}}+e^{\frac{i
   k_x}{2}+\frac{i k_y}{3}}\right)\\
H_{13} &= 0\\
H_{14} &=r_1 \left(-e^{\frac{i k_x}{2}+\frac{i k_y}{3}}\right)+r_2 e^{\frac{i k_y}{3}-\frac{i k_x}{2}}-r_3 e^{-\frac{i k_x}{2}-\frac{2 i k_y}{3}}+r_4 e^{\frac{i
   k_x}{2}-\frac{2 i k_y}{3}}\\
H_{22}&=-\epsilon_0+s_1 \left(e^{-i k_x}+e^{i k_x}\right)+s_2 \left(e^{-i k_y}+e^{i k_y}\right)\\
H_{23}&=r_1 e^{\frac{i k_x}{2}-\frac{i k_y}{3}}-r_2 e^{-\frac{i k_x}{2}-\frac{i k_y}{3}}+r_3 e^{\frac{2 i k_y}{3}-\frac{i k_x}{2}}-r_4 e^{\frac{i k_x}{2}+\frac{2
   i k_y}{3}}\\
H_{24}&= 0\\
H_{33}&=\epsilon_0+s_5 \left(e^{-i k_x}+e^{i k_x}\right)+s_6 \left(e^{-i k_y}+e^{i k_y}\right)\\
H_{34}&=t_4 \left(e^{-\frac{i k_x}{2}-\frac{2 i k_y}{3}}+e^{\frac{i k_x}{2}-\frac{2 i k_y}{3}}\right)-t_3 \left(e^{\frac{i k_y}{3}-\frac{i k_x}{2}}+e^{\frac{i
   k_x}{2}+\frac{i k_y}{3}}\right)\\
H_{44}&=\epsilon_0+s_5 \left(e^{-i k_x}+e^{i k_x}\right)+s_6 \left(e^{-i k_y}+e^{i k_y}\right)
\label{H44}
\end{align}
where $ \epsilon_s=0.477, \epsilon_0=0.103, t_1=0.0214, t_2=-0.00286, t_3=-0.0471, t_4=-0.00429, r_1=0.0143, r_2=0.0214, r_3=0.00714, r_4=0.0143, s_1=0.0143, s_2=0.00429, s_5=-0.0171, s_6=0.0157$.

\paragraph{pmg group:}
The TB Hamiltonian for the $pmg$ lattice with basis orbitals $\ket{A,s}$, $\ket{B,s}$,  $\ket{A,p_y}$ and $\ket{B,p_y}$ is given by
\begin{align}
    H_{11} &= H_{22} = 2 t_0 (\cos k_x+\cos k_y)-\epsilon_o \\
    H_{33} &= H_{44} = 2 r_0 \cos k_x-2 r_0 \cos k_y+\epsilon_o \\
    H_{13} &= -2 i (s_1 \sin k_x-s_2 \sin k_y) \\
    H_{24} &= 2 i (s_1 \sin k_x+s_2 \sin k_y) \\
    H_{12} &= 2 \cos \left(\frac{a k_x}{2}\right) (2 t_2 \cos k_y+t_1) \\
    H_{14} &= H_{23} = -2 i s_3 \sin \left(\frac{a k_x}{2}\right) \\
    H_{34} &= 2 \cos \left(\frac{a k_x}{2}\right) (2 r_2 \cos k_y+r_1) 
\end{align}
For the Dirac point to emerge, the parameters can be chosen to be $\ \epsilon_s=0.228,\ \epsilon_o = 0.1,\ 
t_0 = - 0.00643,\ t_1 = -0.0421 ,\ t_2 = -0.00446 ,\ s_1 =-0.0102,\ s_2 = -0.0102,\ s_3=-0.0357,\ r_0 = -0.0143,\ r_1 = -0.0429,\ r_2 = -0.00893$.

\paragraph{pgg group:} The Hamiltonian in the basis of  $\ket{A,2s}$, $\ket{B,2s}$, $\ket{A,p_y}$ and $\ket{B,p_y}$ orbitals is
\begin{align}
H_{11} &=\epsilon_0+s_1 \left(e^{-i k_x}+e^{i k_x}\right)+s_2 \left(e^{-i k_y}+e^{i k_y}\right)\\
H_{12} &=t_1 \left(1+e^{i k_x}\right) \left(1+e^{i k_y}\right) e^{-\frac{1}{2} i \left(k_x+k_y\right)}\\
H_{13} &=s_3 \left(e^{-i k_x}-e^{i k_x}\right)+s_4 \left(e^{-i k_y}-e^{i k_y}\right)\\
H_{14} &=e^{-\frac{1}{2} i \left(k_x+k_y\right)} \left(r_1 \left(1-e^{i \left(k_x+k_y\right)}\right)+r_2 \left(e^{i
   k_x}-e^{i k_y}\right)\right)\\
H_{22}&=\epsilon_0+s_1 \left(e^{-i k_x}+e^{i k_x}\right)+s_2 \left(e^{-i k_y}+e^{i k_y}\right)\\
H_{23}&=e^{-\frac{1}{2} i \left(k_x+k_y\right)} \left(r_1 \left(e^{i k_x}-e^{i k_y}\right)-r_2 \left(-1+e^{i
   \left(k_x+k_y\right)}\right)\right)\\
H_{24}&=s_7 \left(e^{i k_x}-e^{-i k_x}\right)-s_8 \left(e^{i k_y}-e^{-i k_y}\right)\\
H_{33}&=-\epsilon_0+s_5 \left(e^{-i k_x}+e^{i k_x}\right)+s_6 \left(e^{-i k_y}+e^{i k_y}\right)\\
H_{34}&=t_3 \left(1+e^{i k_x}\right) \left(1+e^{i k_y}\right) e^{-\frac{1}{2} i \left(k_x+k_y\right)}\\
H_{44}&=-\epsilon_0+s_5 \left(e^{-i k_x}+e^{i k_x}\right)+s_6 \left(e^{-i k_y}+e^{i k_y}\right)
\end{align}
with parameters: $\epsilon_0=0.0739, s_1=-0.013, s_2=-0.017,s_3=-0.013, s_4=-0.0043, s_5=0.0043, s_6=0.013, t_1=0.043, t_3=-0.00435, r_1=0.0004, r_2=0.061$.

\paragraph{p4g group:} The Hamiltonian for the $p4g$ case is of the same form as that of the $pgg$ lattice, and its parameters are given by: $ \epsilon_0=0.069, s_1=-0.015, s_2=-0.015,s_3=-0.011, s_4=-0.011, s_5=0.0087, s_6=0.0087, t_1=0.043, t_3=0, r_1=0, r_2=0.061$.

\section{$Z_2$ quantization of 1D Berry phase}
We analyze the topological protection of the Dirac points via $Z_2$ quantization, following methods in Refs.~\cite{yu2011equivalent,lee2013,kariyado2013symmetry}. A 2D lattice is regarded as a family of 1D systems indexed by a $2\pi$-periodic parameter $k_p$. Within each 1D system, there also exists a $2\pi$-periodic momentum $k_o$. For the case of our four band model Hamiltonian, the (non-abelian) Berry phase is defined as
\begin{equation}
    \gamma(k_p) = -i\sum_{n\in\text{filled}}\int_{-\pi}^{\pi}dk_o\bra{\varphi_{n,\bf{k}}}\partial_{k_o}\ket{\varphi_{n,\bf{k}}},
\end{equation}
where $\ket{\varphi_{n,\bf{k}}}$ are the eigenstates of the Hamiltonian and $k_{o},k_p\in \{k_x,k_y\}$. 

To simplify notation, we define filled state vectors as $\Psi=(\ket{\varphi_1}, \ket{\varphi_2})$, with $\bf{k}$ indices suppressed. The Berry phase computation is done on a lattice~\cite{Hirano2008prb}, such that
\begin{equation}
\begin{split}
    \gamma(k_p) &= -\Im\Tr\oint_C dk_o \Psi^{\dagger}\partial_{k_o}\Psi\\
    &\approx -\Im \Tr\sum_m \Delta k_o \Psi^{\dagger}_{m\Delta k_o}\partial_{k_o}\Psi_{m\Delta k_o}\\
    &\approx -\Im \Tr\log[\Pi_m\Psi^{\dagger}_{m\Delta k_o}\Psi_{(m+1)\Delta k_o}],
\end{split}
\end{equation}
where $C$ is a closed path in the direction of $k_o$ and we have used $\Psi^{\dagger}_{m\Delta k_o} \Psi_{(m+1)\Delta k_o}\approx \mathbb{I}_2+\Delta k_o \Psi^{\dagger}_{m\Delta k_o} \partial_{k_o}\Psi_{m\Delta k_o}\approx \text{exp}(\Delta k_o \Psi^{\dagger}_{m\Delta k_o} \partial_{k_o}\Psi_{m\Delta k_o})$ in the last step. We recognize the operator in the argument of the logarithm as the Wilson loop operator, which we denote as $\mathcal{W}_{k_o,k_p}$, from which a Wannier Hamiltonian can be defined via $\mathcal{W}_{k_o,k_p}:=\exp[iH_{\mathcal{W}}(k_p)]$\cite{Benalcazar2016arXiv}, which also qualitatively controls the evolution of the entanglement spectrum\cite{lee2015free}. The 1D Berry phase may then be interpreted as the sum of eigenvalues, $2\pi\nu^i$, of the Wannier Hamiltonian, where $\nu^{1,2}$ are known as the Wannier centers. The periodic nature of the Wannier centers, due to the unitarity of Wilson loops, is key to understanding the $Z_2$ quantization of the 1D Berry phase. 

Since the Wilson loop operator describes the winding number of a mapping from a 1D loop, for it to be quantized, it is essential to impose certain symmetries that restricts the state space to another 1D manifold. An example of such a symmetry operation is the mirror operator $M_o$, where $M_o(k_o,k_p)=(-k_o,k_p)$. Since this mirror essentially reverses the direction of the Wilson loop, i.e. $\mathcal{W}_{k_o,k_p}\rightarrow\mathcal{W}^{\dagger}_{k_o,k_p}$ and the Wannier centers are independent, we arrive at the constraint
\begin{equation}
    \left\{e^{i2\pi\nu^i(k_p)}\right\}=\left\{e^{-i2\pi\nu^i(k_p)}\right\}.
\end{equation}
One important implication of this constraint is that the Wannier centers must now  come in pairs of the forms $(-\nu,\nu)$ or $(0,1/2)$, the sum of which is quantized to $0$ or $1/2$ respectively. This argument applies in the same way when the system has glide mirror symmetry, as in our case.

Evidently, the 1D Berry phase is classified with a $Z_2$ index, and by continuously tuning the parameter $k_p$, one may observe a phase transition at gap closing points, which correspond to (projected) Dirac points in our 2D system. In addition, the zero modes in the 1D topological $Z_2$ insulators will compose to edge states in 2D. This observation allows us to efficiently study the emergence of edge states connecting the Dirac points in the four nonsymmorphic groups. We demonstrate this approach by computing the $Z_2$ quantized 1D Berry phase for all four groups, as presented in the Fig.~\ref{fig:Z2}.

Note that in order to calculate the $Z_2$ index and band structure of edge states, the tight-binding Hamiltonian elements should have $2\pi$ periodicity in the momenta. This can be achieved with a gauge transformation given by $H_{12}^B = H_{12} e^{i\beta}$, $H_{14}^B = H_{14} e^{i\beta}$, $H_{23}^B = H_{23} e^{-i\beta}$, $H_{34}^B = H_{34} e^{i\beta}$, with the rest elements invariant. For the $pg$ group, $\beta_{pg} = {\frac{-k_x}{2}+\frac{-k_y}{3}}$. For $pmg$, $p4g$ and $pgg$ groups, $\beta_{pmg} = {\frac{-k_x}{2}}$, $\beta_{p4g} = \beta_{pgg} = {\frac{-k_x}{2}+\frac{-k_y}{2}}$.

In Fig.~\ref{fig:edge}, we take illustrate the edge states of our $pmg$ and $pgg$ systems. Evidently, edge states appear in both the simulation results and TB model (with open boundary condition) at exactly the same momenta as the topologically nontrivial regions appearing in Fig.~\ref{fig:Z2}. In the photonic crystal simulations, the open boundary condition is implemented by adjoining the crystal with a trivial lattice such as a lattice of cylinders with square cross sections, whose parameters are given in Table.~\ref{tab:tri}.


\begin{table}[ht!]
\centering
\caption{Trivial square lattice parameters for implementing the open boundary in the COMSOL simulation.}
\label{tab:tri}   
\begin{tabular}{ccccc}
\hline 
case & $r$ & position & $\varepsilon_r$ &gap range \\
\hline
$pmg$ lattice boundary& $0.3a$ &  (0,0) & 27 & 0.40~0.52\\
$pgg$ lattice boundary& $0.22a$&(-a/4,-a/4) (a/4,a/4) & 15 & 0.78~0.90\\ 
\hline 
\end{tabular} 
\end{table}
\begin{figure}[t]
\subfloat[]{\label{fig:Z2pg}%
\includegraphics[width=0.22\textwidth]{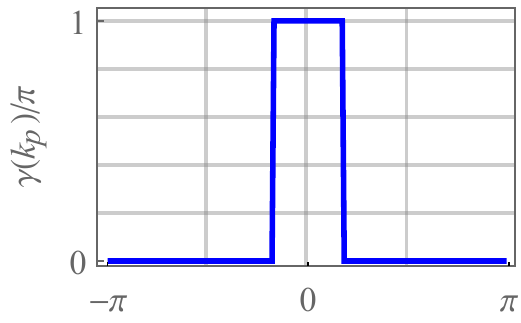}%
}
\subfloat[]{\label{fig:Z2pmg}%
  \includegraphics[width=0.22\textwidth]{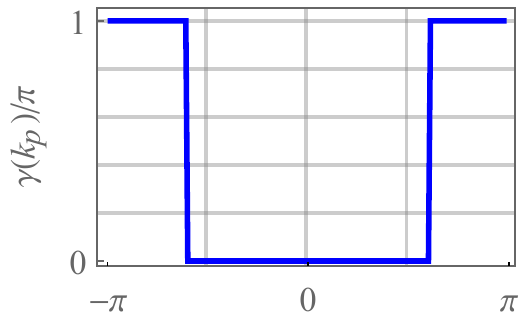}%
}
\subfloat[]{\label{fig:Z2p4g}%
\includegraphics[width=0.22\textwidth]{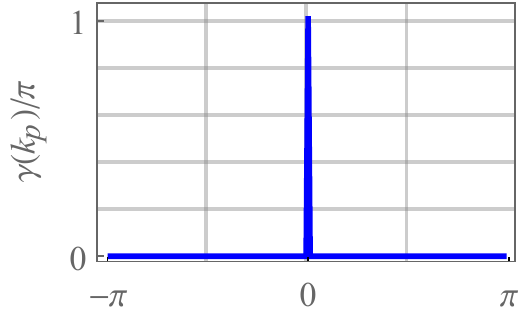}%
}
\subfloat[]{\label{fig:Z2pgg}%
\includegraphics[width=0.22\textwidth]{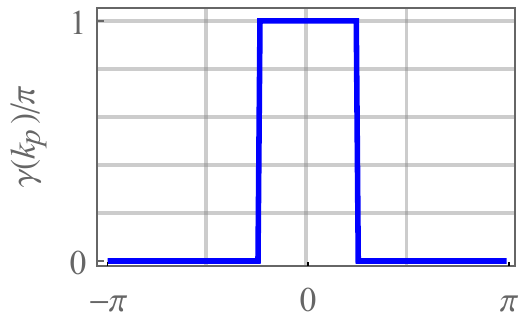}%
}
\caption{$\gamma(k_p)$ for the$pg$, $pmg$, $p4g$ and $pgg$ lattices, which are quantized to $0$ or $1$. Protected Dirac points exist at its discontinuities, where the gap has to close. In the $p4g$ case, the Dirac points coalesce, making the nontrivial $Z_2$ region into a sharp spike.
		}
\label{fig:Z2}
\end{figure}
\begin{figure}[t]
\subfloat[]{\label{fig:edgepmgsimu}%
\includegraphics[width=0.22\textwidth]{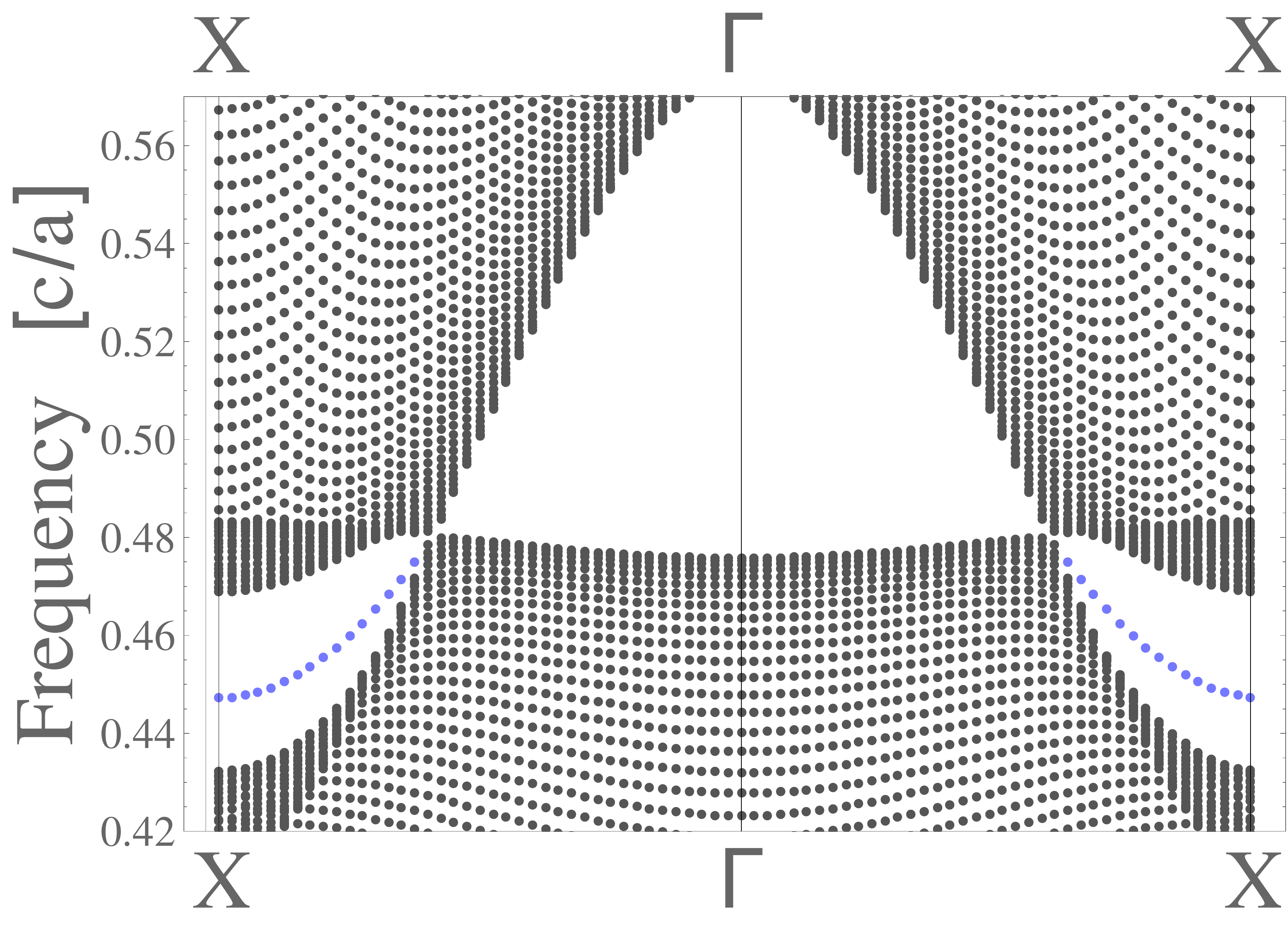}%
}
\subfloat[]{\label{fig:edgepmgmodel}%
  \includegraphics[width=0.20\textwidth]{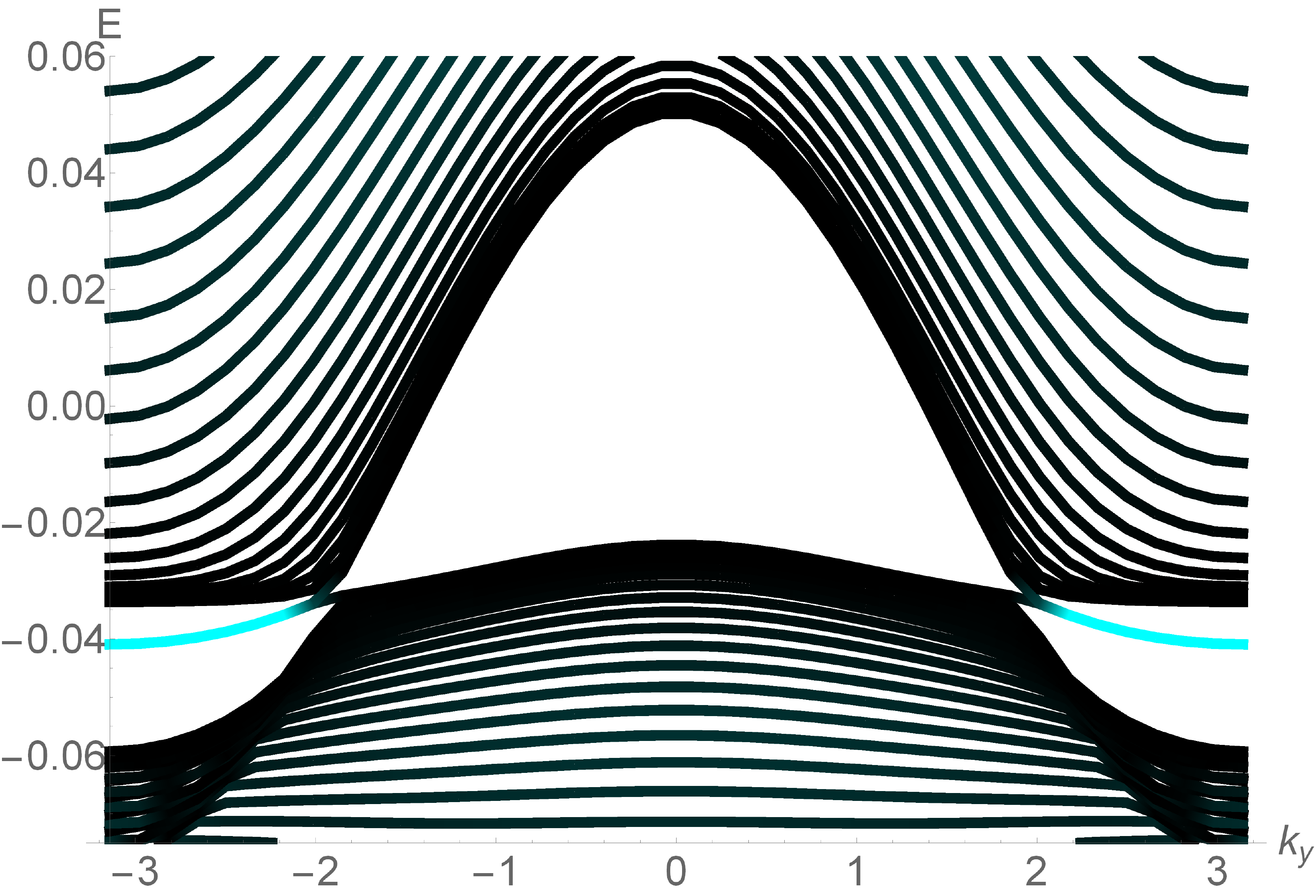}%
}
\subfloat[]{\label{fig:edgepggsimu}%
\includegraphics[width=0.22\textwidth]{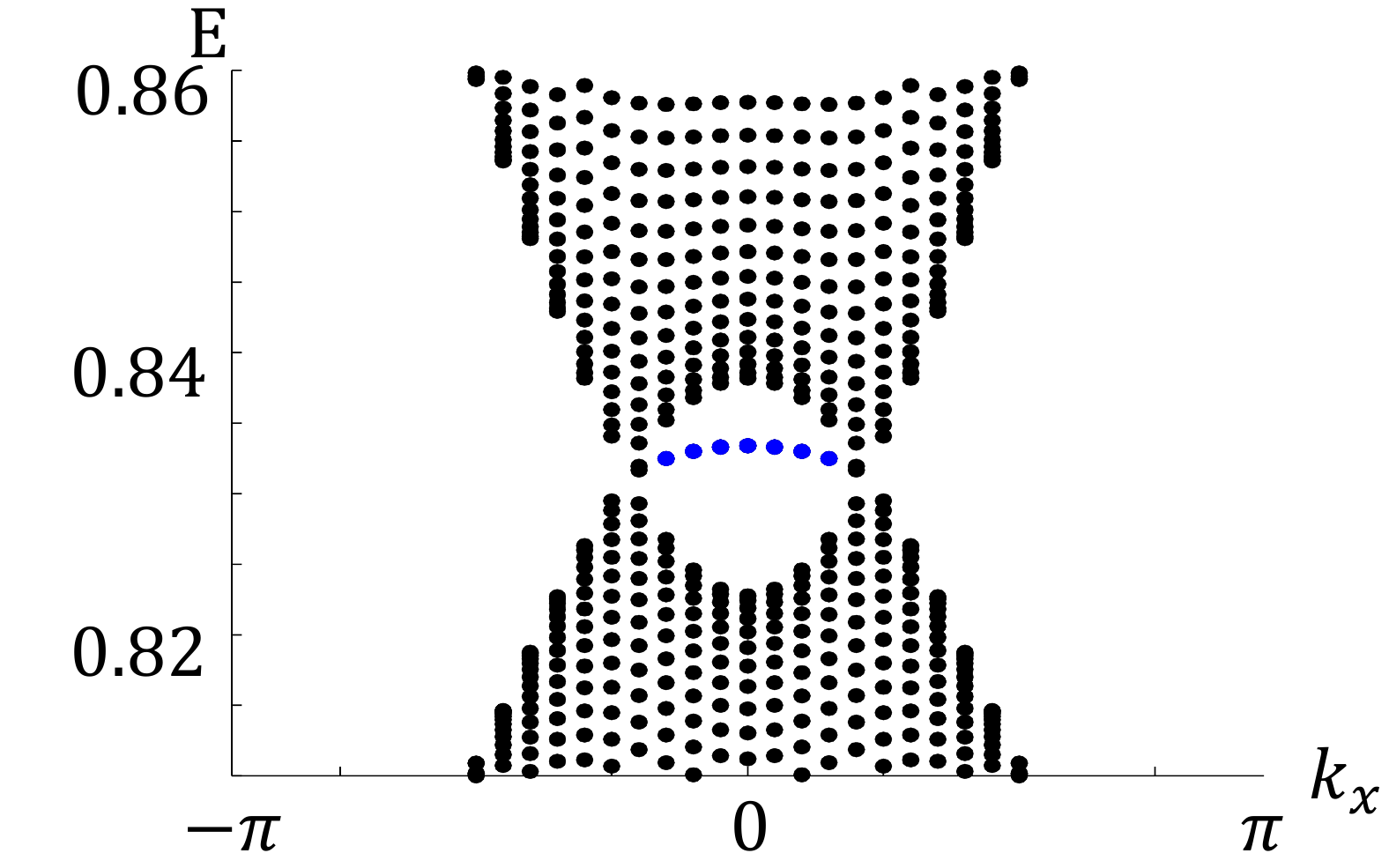}%
}
\subfloat[]{\label{fig:edgepggmodel}%
  \includegraphics[width=0.20\textwidth]{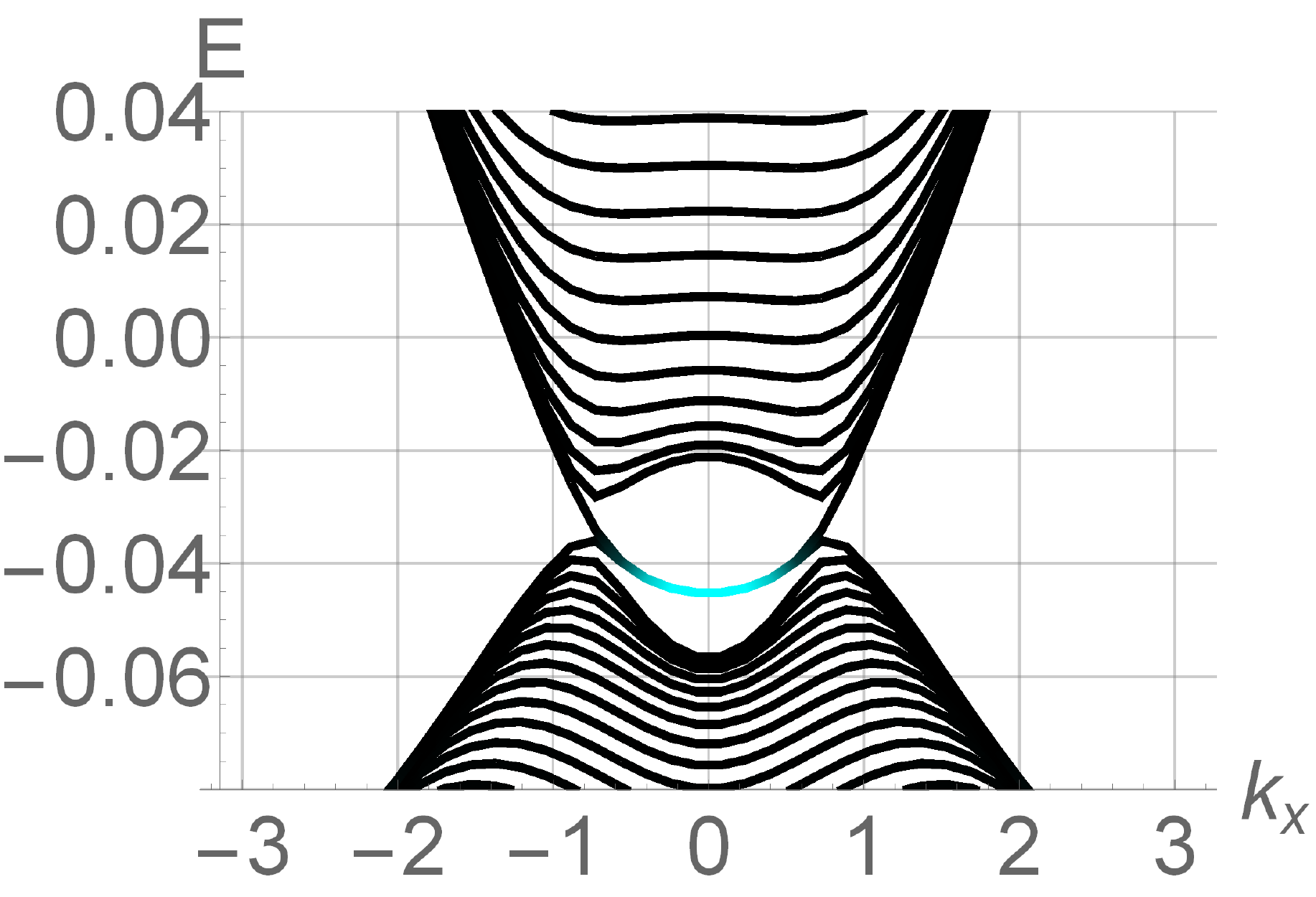}%
}
    \caption{Edge states from simulation results of the (a)$pmg$ lattice and (c)$pgg$ lattice, which agree well with those from the (b) $pmg$ and (d) $pgg$ tight-binding models with open boundary conditions. Indeed, the edge states exists across $Z_2$ nontrivial regions in Fig. \ref{fig:Z2}, which are terminated by point degeneracies.
		}
\label{fig:edge}
\end{figure}

\section{Perturbation theory and vortex structure}
A 4$\times$4 Hamiltonian $H$ can be reduced to a $2\times 2$ 
Hamiltonian by perturbing around the photon frequency $\omega$. Define $P$ and $Q=\mathbb{I}-P$ respectively as the projector onto the desired/truncated $2\times 2$ subspaces. The reduced Hamiltonian $H_{red}$ is given by
\begin{equation}
H_{red}= PHP+PHQ[Q(H-\omega)Q]^{-1}QHP+...
\end{equation}
with terms up to second order shown. 
Performing this perturbation on the nonsymmorphic Hamiltonians near the frequencies of the point degeneracies, we obtain the exact form of the $\mathbf{h}$ vector. Since nontrivial topology only appears when $\mathbf{h}$ vector is constrained to 2-component in 2D systems, we can plot the change of $\mathbf{h}$ near the Dirac points for the $pmg,pgg$ and $p4g$ lattices. Noting that the magnitude of the $\mathbf{h}$ is an indicator of band gap, each Dirac point gives a vortex core as shown in Fig.~\ref{fig:hvector}.

\section{Detailed argument for line degeneracy due to $pg$ symmetry}
The 2$\times$2 TB Hamiltonian is written in the basis $\ket{A,\mathbf{k}}$ and $\ket{B,\mathbf{k}}$:
\begin{equation}
    H_{pg} = \left(
\begin{array}{cc}
 H_A & H_{AB}\\
H_{BA}&  H_B \\
\end{array}
\right).
\end{equation}
whose eigenvalues yield $\omega^2_{n,\mathbf{k}}$. We show that there is a nodal line (line of degenerate states) along the path $X$-$M$, where $k_x=\frac{\pi}{a}$. To see how this degeneracy arises due to symmetry under $g_y=\{m_y|\tau_x\}$, we first analyze what it constraints. Trivially, it implies that the distance between two neighboring elliptical cylinders is always $a/2$ in the x-direction. Hence the orbital overlaps in the TB model must always satisfy $J_{n,m}=J_{-n,m}$, where $n,m$ label the horizontal/vertical coordinates. As such, $H_{AB}(\pi/a,k_y)$ always evaluates to
\begin{equation}
    \begin{split}
        H_{AB}\left(\frac{\pi}{a},k_y\right)&= \sum_{m=-\infty}^{+\infty}\sum_{n=0}^{+\infty} J_{n,m}e^{i k_x a(\frac{1}{2}+n)+i k_y a (\frac{1}{2}+m)}\\
        &+J_{-n,m}e^{i k_x a(-\frac{1}{2}-n)+i k_y a (\frac{1}{2}+m)}|_{k_x=\frac{\pi}{a}}\\
        &= \sum_{m=-\infty}^{+\infty}\sum_{n=0}^{+\infty}2i J_{n,m} e^{i k_y a (\frac{1}{2}+m)}\sin(n \pi )\\
        &= 0.
    \end{split}
\end{equation}
In a nutshell, the projective phase factor evaluates to $-1$ along $\textbf{k}=(\pm \frac{\pi}{a},k_y)$, which forces the off-diagonal Hamiltonian matrix elements to vanish: $H_{BA}(-\frac{\pi}{a},k_y)=-H_{AB}(\frac{\pi}{a},k_y)=0$. 
Constrained by $g_y$, the remaining matrix elements $H_{AA}(\pi/a,k_y)$ and $H_{BB}(\pi/a,k_y)$ must be identical. Hence the double degeneracy of their two eigenstates $\ket{A,s,\mathbf{k}}$ and $\ket{B,s,\mathbf{k}}$.

\section{Derivation of anomalous refraction angles}
Since the frequency $\omega$ and tangential component $k_y$ of the wave vector are conserved during refraction, we can easily determine the group velocities of the refracted beams through their normals to the isofrequency contours of a type-II Dirac cone. This can be performed graphically as shown in Fig.~\ref{fig:typeIIrefract}, and we find two refracted beams that have opposite refraction angles. 
Explicitly, the group velocities for the Dirac point dispersion $\delta\omega=\omega-\omega_0=\eta v_x \delta k_x \pm \sqrt{v_x^2 \delta k_x^2+v_y^2\delta k_y^2}$ are given by
\begin{equation}
    v^g_y = \frac{\pm v_y^2\delta k_y}{\sqrt{v_x^2 \delta k_x^2+v_y^2\delta k_y^2}},\  
    v^g_x = \eta v_x \pm \frac{v_x^2 \delta k_x}{\sqrt{v_x^2 \delta k_x^2+v_y^2\delta k_y^2}},
\end{equation}
which yields refraction angles
\begin{equation}
\begin{split}
    \phi^{\pm} &= \tan^{-1}\frac{v^g_y}{v^g_x}\\
		&= \tan^{-1} \frac{\pm v_y^2\delta k_y/\sqrt{v_x^2 \delta k_x^2+v_y^2\delta k_y^2}}{\eta v_x \pm v_x^2 \delta k_x/\sqrt{v_x^2 \delta k_x^2+v_y^2\delta k_y^2}} \\
    &= \pm \tan^{-1} \frac{ v_y^2 |\delta k_y|}{\eta v_x(\delta\omega-\eta v_x \delta k_x) + v_x^2 \delta k_x},
\end{split}
\end{equation}
where
\begin{equation}
    |\delta k_y| = \sqrt{(\delta\omega-\eta v_x \delta k_x)^2-v_x^2 \delta k_x^2}/v_y
\end{equation}
and $\delta k_x = \frac{\omega}{c}\sin\theta$, with $\theta$ the incident angle. Exactly at the frequency of the Dirac point, $\delta \omega=0$ and $|\delta k_y| = |\delta k_x|\frac{v_x}{v_y}\sqrt{\eta^2-1}$. This yields $\phi^\mp = \pm \tan^{-1}\left[\frac{v_y}{v_x}\frac1{\sqrt{\eta^2-1}}\right]$,  which clearly implies the existence of anomalous refraction only if $|\eta|>1$, i.e. if the Dirac point is of type II.

Since we physically require $v^g_x>0$, the sign of $\eta$ is fixed. But due to time reversal symmetry, the two Dirac points come with opposite signs of $\eta$. Hence only one tilted (type-II) Dirac point can induce anomalous refraction. This property may be used to design filters in valleytronics applications.

As a rough illustration, if the incident beams were come from air with angle $\theta = 6^{\circ}$, the angles of refraction will be $\phi_1  \approx -\phi_2 \approx 50^{\circ}$ according to the PhC parameters given.
\end{widetext}

\end{document}